\documentstyle[prb,twocolumn,aps]{revtex}
\begin{document}
\draft
\begin{title}
{Green's and spectral functions of the small Fr\"ohlich polaron}
\end{title}
\author{A.S. Alexandrov and C. Sricheewin}
\address
{Department of Physics, Loughborough University, Loughborough LE11
3TU, U.K.}

\maketitle
\begin{abstract}
According to  recent Quantum Monte Carlo simulations  the small polaron
theory  is practically
exact in a wide range of the long-range (Fr\"ohlich)
electron-phonon coupling and adiabatic ratio. We apply  the Lang-Firsov transformation 
 to convert the strong-coupling term 
in the Hamiltonian into the form of an  effective hopping integral and
derive  the single-particle Green's function describing  propagation
of  the small Fr\"ohlich polaron. One
and two dimensional spectral functions are studied by expanding
 the Green's function perturbatively.
Numerical calculations of the spectral functions are
produced. Remarkably, the coherent spectral weight 
 ($Z$) and effective mass ($Z'$) renormalisation exponents are found
to be different with $Z'>>Z$, which can  explain a small coherent spectral weight and   a relatively
moderate mass enhancement in oxides.

\end{abstract}
\narrowtext

\centerline {\bf 1. Introduction}

The problem of a fermion on a lattice  coupled with the bosonic
field of lattice vibrations has an exact solution in terms of the
coherent (Glauber) states in the extreme strong-coupling limit,
$\lambda =\infty$ for
any type of  electron-phonon interaction conserving the on-site
occupation numbers of fermions \cite{mah,alemot}. For the 
intermediate coupling   the $1/\lambda$ perturbation diagrammatic
technique has been developed both 
for a single \cite{lan} and multi-polaron systems \cite{ale}. 
 The expansion parameter  actually is $1/2z\lambda^{2}$ \cite{lan,eag,gog,ale}, so the analytical 
 perturbation theory might have  a wider region of applicability than one can 
 expect from a naive variational
 estimate ($z$ is the lattice coordination number).  However,  it is not clear
   how fast the expansion converges.   The exact numerical diagonalisation  of  vibrating clusters,
variational calculations 
     \cite{kon,alekab,bis,feh,mar,feh2,rom,mag,jec,tru,wag}, dynamical
mean-field approach in infinite dimensions \cite{zey}, and  Quantum-Monte-Carlo simulations 
     \cite{rae,kor,alekor,kor2}  
  revealed that   the ground 
 state energy (the polaron binding energy $E_{p}$)  is  not very sensitive to the 
 parameters. On the contrary,  the effective 
 mass, the bandwidth and the shape of  polaron density of states  strongly depend  on  polaron size and 
 adiabatic ratio in  case of a short-range (Holstein) interaction. 
 In particular,  numerical diagonalisation of the two-site-one-electron Holstein model
 in the adiabatic $\omega_{0}/t<1$ as well as 
in the nonadiabatic $\omega_{0}/t>1$ 
regimes  shows
 that perturbation theory is almost exact 
in the nonadiabatic regime  $for$ $all$
$values$ of the coupling constant. However,   there is no agreement 
in the adiabatic region,  where the first order perturbation 
expression $overestimates$  the polaron mass by a few orders of 
magnitude in the intermediate coupling regime \cite{alekab}. Here
$\omega_{0}$ is the characteristic phonon frequency and $t$ is the 
nearest-neighbour hopping integral so that the dimensionless coupling constant
is $\lambda= E_{p}/(zt)$. A much lower effective mass of the adiabatic small polaron in the 
  intermediate coupling regime compared with that estimated by 
 first order perturbation theory  is a result of 
 poor convergence of the perturbation expansion owing to
 the appearance of the familiar double-well potential 
\cite{hol} in  the adiabatic limit. The tunnelling probability is 
 extremely sensitive to the shape of this potential.

It has
 also been  understood \cite{alekor} that the range of applicability  of
 the analytical theory \cite{lan,ale}
strongly depends on the radius of  interaction.
While  the analytical approach is  applicable only if $\omega_{0} \geq
t$ for   short-range 
 interaction,
 the theory appears almost exact in a substantially wider region of the
 parameters for a long-range (Fr\"ohlich) interaction. The exact effective mass
of {\it both} small $and$ large 
Fr\"ohlich polarons, calculated  with  
the continuous-time path-integral Quantum Monte Carlo
(QMC) algorithm,
$m^{\ast}(\lambda)$ is well fitted by
a single exponent \cite{alekor,kor2}. As an example, $e^{0.73 \lambda}$ for $ \omega_{0} = t$
and  $e^{1.4 \lambda}$ for $\omega_{0} = 0.5 \, t$ describe
$m^{\ast}(\lambda)$
in a one-dimensional lattice. 
 The numerical exponents
 are remarkably close to those obtained from  the
Lang-Firsov transformation, $e^{0.78\lambda}$ and $e^{1.56\lambda}$,
 respectively. Hence, in the case of the Fr\"ohlich interaction
 the transformation is perfectly accurate already in the first order
of  $1/\lambda$ expansion 
 even in the adiabatic regime, 
 $\omega_{0}/t \leq 1$ for $any$ coupling strength.

In this paper we use this  result to calculate the 
single-particle Green's function of a fermion on a lattice coupled with the bosonic
field via the long-range Fr\"ohlich interaction.

\vspace{0.5cm}

\centerline {\bf 2. Green's functions of the small Fr\"ohlich polaron }

The classical approach to the small polaron problem is based on the 
canonical displacement (Lang-Firsov) transformation of the 
electron-phonon Hamiltonian \cite{lan} allowing for the summation of all 
diagrams including the vertex corrections,
\begin{eqnarray}
 H &=& \sum_{i,j}t({\bf m-n})\delta_{s,s'}
 c^{\dagger}_{i}c_{j}+\sum_{{\bf q}, 
 i}\omega_{\bf q}\hat{n}_{i}\left[u_{i}({\bf q})d_{\bf q} + H.c.\right] \cr 
 &+& 
 \sum_{\bf q}\omega_{\bf q}(d_{\bf q}^{\dagger}d_{\bf q}+1/2)
 \end{eqnarray}
 with the bare hopping integral $t({\bf m })$
and the matrix element of the electron-phonon interaction
 \begin{equation}
 u_{i}({\bf q})={1\over{\sqrt{2N}}}\gamma({\bf q})e^{i{\bf q\cdot 
 m}}.
 \end{equation}
 Here $i=({\bf m},s)$, $j=({\bf n},s')$, include site ${\bf m}$ and
spin $s$, 
$\hat{n}_{i}=c^{\dagger}_{i}c_{i}$, and $c_{i},d_{\bf q}$ are  
electron (hole) 
and phonon operators, respectively.

The canonical transformation $e^{S}$  diagonalising the Hamiltonian
 in  the $\lambda=\infty$ limit is
\begin{equation}
\tilde {H}=e^{S} H e^{-S},
\end{equation}
where
\begin{equation}
S=\sum_{{\bf q}, 
 i}\hat{n}_{i}\left[u_{i}({\bf q})d_{\bf q} -H.c.\right].
 \end{equation}
  The electron operator transforms as
 \begin{equation}
 \tilde{c}_{i}=c_{i} exp\left(-\sum_{\bf q}u_{i}({\bf q})d_{\bf 
 q}-H.c.\right)\label{eq:100}
 \end{equation}
 and  the phonon one as:
 \begin{equation}
 \tilde{d}_{\bf q}= d_{\bf q}+\sum_{i}\hat{n}_{i}u^{*}_{i}({\bf 
 q})
 \end{equation}
It follows from Eq.(6)  that the Lang-Firsov transformation 
 shifts  ions to new equilibrium positions. In a more general sense it 
 changes the boson vacuum. As a result,
 \begin{eqnarray}
 \tilde{H}&=&\sum_{i,j}\hat{\sigma}_{ij}c^{\dagger}_{i}c_{j}
 -E_{p}\sum_{i}\hat{n}_{i}  \cr
 &+&\sum_{\bf 
 q}\omega_{\bf q}(d_{\bf q}^{\dagger}d_{\bf q}
 +1/2)+{1\over{2}}\sum_{i\neq j}v_{ij}\hat{n}_{i}\hat{n}_{j},
 \end{eqnarray}
where
\begin{equation}
\hat{\sigma}_{ij}=t({\bf m-n})\delta_{s,s'}\exp\left(\sum_{\bf q}[u_{i}({\bf q})-u_{j}({\bf q})]d_{\bf q} -H.c.\right)
 \end{equation}
 is a renormalised hopping integral depending on the phonon variables,
\begin{equation}
E_{p}= {1\over{2N}}\sum_{\bf q}|\gamma({\bf q})|^{2}\omega_{\bf 
 q}
\end{equation}
the polaron level shift (binding energy),  and
 \begin{equation}
 v_{ij}= -{1\over{N}}\sum_{\bf q}|\gamma({\bf q})|^{2}\omega_{\bf 
 q}\cos[{\bf q \cdot (m-n})]
 \end{equation}
   the attractive interaction of polarons 
owing to   lattice deformation.

We consider a single-particle problem when the last
term in Eq.(7) is absent. The QMC result \cite{alekor} allows us to average
the first hopping term with respect to phonons no matter what the
value of  the Fr\"ohlich interaction is. Then we can  apply  the canonical
trasformation together with the averaged Hamiltonian to derive the 
single-particle Green's function (GF). The interaction is
completely  integrated out from  the averaged Hamiltonian,
\begin{equation}
\tilde{H}= H_{p}+H_{ph}
\end{equation}
where the 'free' polaron part is given by
\begin{equation}
H_{p}=\sum_{\bf k}\xi({\bf k})c^{\dagger}_{\bf k} c_{\bf k},
\end{equation}
(from now we omit the spin index), and the free phonon part is
\begin{equation}
H_{ph}=\sum_{\bf 
 q}\omega_{\bf q}(d_{\bf q}^{\dagger}d_{\bf q}
 +1/2).
\end{equation}
Here $\xi({\bf k})=Z'E({\bf k})-\mu$ is the renormalised polaron-band
dispersion with the chemical potential $\mu$, which includes 
polaron binding energy ($-E_{p}$) \cite{ref}, $E({\bf k})= \sum_{\bf m}t({\bf
m})exp(-i{\bf k \cdot m})$ is the bare disperison in a rigid lattice, and the
mass-renormalisation exponent is
\begin{equation}  
Z'={\sum_{\bf m}t({\bf
m})e^{-g^2({\bf m})}\exp(-i{\bf k \cdot m})\over{\sum_{\bf m}t({\bf
m})\exp(-i{\bf k \cdot m})}}
\end{equation}
with 
\begin{equation}
g^2({\bf m})=\sum_{\bf q} |\gamma({\bf q})|^{2} [1-cos ({\bf q \cdot
 a})].
\end{equation}
 Quite generally
 one finds $Z'=exp(-\gamma E_{p}/\omega)$, where the numerical coefficient
 \begin{equation}
 \gamma= \sum_{\bf q} |\gamma({\bf q})|^{2} [1-\cos ({\bf q \cdot a})]/
\sum_{\bf q} |\gamma({\bf q})|^{2},
\end{equation}
might be  as small as
 $0.4$\cite{alekor}
and
even smaller in the
cuprates with  nearest neighbour oxygen-oxygen distance less than the
lattice constant,
$\gamma \simeq 0.2$ \cite{ale2,alecom}.

Applying  the Lang-Firsov canonical
transformation the Fourier component of the retarded GF,
\begin{equation}
 G_{R}({\bf
k},\omega)=-i\sum_{\bf m}\int_{0}^{\infty} e^{-i({\bf k
\cdot m}-\omega t)}
\langle
0|c_{0}(t)c^{\dagger}_{\bf m}(0)|0\rangle dt
\end{equation}
  is expressed as a convolution of the
Fourier components of the
coherent retarded polaron GF, $G_{p}( {\bf n}, t)$, and  the multiphonon correlation function
$\sigma({\bf n},t)$:
 \begin{equation}
G_{R}({\bf k}, \omega)={1\over{2\pi}} \sum_{\bf m}e^{-i{\bf k \cdot
n}}\int _{-\infty}^{\infty} d\omega' G_{p}({\bf m},\omega')\sigma({\bf
m}, \omega-\omega').  
\end{equation}
Here
\begin{equation}
G_{p}({\bf m},\omega)=- i \int_{0}^{\infty} dt e ^{i\omega t}
\langle \tilde{0}| e^{iH_{p}t} c_{0} e^{-iH_{p}t}c^{\dagger}_{\bf
m}|\tilde{0} \rangle, 
\end{equation}
and
\begin{equation}
\sigma({\bf m}, \omega)= \int_{0}^{\infty} dt e^{i\omega t}\langle \tilde{0}| e^{iH_{ph}t} X_{0} e^{-iH_{ph}t}X^{\dagger}_{\bf
m}|\tilde{0} \rangle
\end{equation}
with $X_{\bf m}= \exp \left( \sum_{\bf q} u_{i}({\bf q})d_{\bf
q}-H.c.\right)$,
and $|\tilde{0}\rangle$ the ground state of $\tilde{H}$. 

Straightforward calculations \cite{mah,alemot} yield
\begin{equation}
G_{p}({\bf m},\omega)={1\over{N}}\sum_{\bf k'} {e^{i\bf k' \cdot
m}\over{\omega-\xi({\bf k'})+i\delta}},
\end{equation}
and
\begin{eqnarray}
\sigma({\bf m}, \omega)=i Z
\sum_{l=0}^{\infty}{1\over{l!(\omega-\omega({\bf q})l+i\delta)}}\cr
\left({1\over{2N}}\sum_{\bf q}|\gamma({\bf q})|^{2}\exp(i\bf q \cdot
m)\right)^l.
\end{eqnarray}
Here
\begin{equation}
Z=\exp (-\sum_{\bf q}|\gamma({\bf q})|^{2}).
\end{equation}  
In the following we consider  dispersionless
phonons, $\omega_{\bf q}=\omega_{0}$, and  the Fr\"ohlich interaction
with $\gamma({\bf q}) \sim 1/q$. The convolution of Eq.(19)
and Eq.(20) yields
\begin{equation}
G_{R}({\bf k}, \omega)= \sum_{l=0}^{\infty} G_{R}^{(l)}({\bf k},
\omega),
\end{equation}
where 
\begin{eqnarray}
&G&_{R}^{(l)}({\bf k},
\omega)= \cr &Z&\sum_{{\bf q}_{1},...{\bf q}_{l}}{|\gamma({\bf
q}_{1})|^{2}\times
|\gamma({\bf q}_{2})|^{2}\times ...\times|\gamma({\bf
q}_{l})|^{2}\over{(2N)^ll!
(\omega-\omega_{0} l-\xi({\bf k+q}_{1}+...{\bf q}_{l})+i\delta)}}
\end{eqnarray}
with $\delta=+0$.

Obviously, Eq.(22) is in the form of a perturbative multiphonon  expansion.
A term with an index $l$ corresponds to a transition from  the initial state ${\bf k}$ of the polaron
band to the final state ${\bf k+q_{1}+...q_{l}}$ with the   emission
of $l$ optical phonons \cite{ref2}.

\vspace{0.5cm}

\centerline {\bf 3. Spectral functions of the small Fr\"ohlich polaron }

The  Green's function of a polaronic carrier, Eq.(22) comprises
 two different contributions. The first ($l=0$) coherent ${\bf k}$-dependent
 term arises from the
polaron band tunneling,
\begin{equation}
G_{R}^{(0)}={Z\over{\omega-\xi({\bf k})+i\delta}}.
\end{equation}
  The spectral weight of the coherent part is
strongly (exponentially) suppressed as $Z= \exp(-E_{p}/\omega_{0})$ while the effective
mass might only be slightly  enhanced, $\xi_{\bf k}=Z' E_{\bf k}-\mu$,
because  $Z<<Z'<1$ in the case of the Fr\"ohlich interaction.

The second incoherent  phonon-assisted contribution with $l\geq 1$
describes the excitations accompanied by  emission  of
phonons.
 We believe that this term  is responsible for the
 background in  optical conductivity and in 
photoemission spectra of cuprates and manganites. We notice that its
spectral density  spreads over a wide energy range
of about  twice the polaron level shift $E_{p}$. On the
contrary the coherent term shows an angular dependence in the energy
window of the
order of the polaron bandwidth, $2Z'zt$.
Interestingly, there is   some ${\bf k}$ dependence of the $incoherent$
background as well, if   the matrix
element of  electron-phonon interaction depends on $q$ (see also
Ref.\cite{kay}). To illustrate this point we calculate the 
single-phonon contribution ($l=1$) to the spectral function 
\begin{equation}
A({\bf
k},\omega)\equiv -(1/\pi) \Im G_{R}({\bf k}, \omega)=
\sum_{l=0}^{\infty}A^{(l)}({\bf k},\omega).
\end{equation}

The coherent  part ($l=0$) of the  spectral function is a $\delta$-function in
agreement with the well-established  fact ( see \cite{alemot,mah} and
references therein) that small
polarons exist in the Bloch states at zero temperature no matter which
values the parameters of the system take. The single-phonon
contribution to the incoherent background is given by
\begin{equation}
A^{(1)}({\bf k},\omega) \sim \int_{BZ}d{\bf q} q^{-2}\delta
(\omega-\omega_{0}- \xi({\bf k+q}),
\end{equation}
where the interval  of 
integration  (the Brillouin zone, BZ) is determined by  lattice
constants $a,b,c$. We calculate this integral for  one-dimensional (1D),
$\xi({\bf k})= 2\tilde{t} \cos (k_{x}a)$ and two-dimensional (2D), $\xi({\bf k})=2\tilde{t}
\cos (k_{x}a)+2\tilde{t}' \cos(k_{y}b)$ polaron bands in the tight-binding
approximation with the (renormalised) nearest neighbour hopping
integrals, $\tilde{t}= Z't$, $\tilde{t}'=Z't'$. 

The 1D single-phonon spectral function is reduced to
\begin{eqnarray}
A^{(1)}({\bf k},\omega) \sim (&1&-\tilde{\omega}^2)^{-1/2} \cr 
\int_{0}^{\pi/3} dz [&f&(z,\omega,k_{x})+f(z,\omega,-k_{x})],
\end{eqnarray}
where
\begin{eqnarray}
f(z,\omega,k_{x})= [z^2 &+&(k_{x}-cos^{-1}
\tilde{\omega})^2]^{-1/2}   \cr   \tan^{-1}(&\pi& [z^2+(k_{x}-cos^{-1}
\tilde{\omega})^2]^{-1/2}).
\end{eqnarray}
Here and further we take $a=b=1$ and $c=3$, and
$\tilde{t'}=\tilde{t}/4$. Its spectral and angular dependence are
shown in Fig.1. Apart from two
nondispersive 1D
van-Hove  singularities (vHs) at $\tilde{\omega} \equiv
(\omega-\omega_{0})/2\tilde{t}=\pm 1$ there is an interesting dispersive peak at
$\tilde{\omega}= cosk_{x}$, which is due entirely to the
long-range character of the Fr\"ohlich interaction. Indeed,
approximating the Brillouin zone by a cylinder along $x$, one readely
obtains a logarithmic singularity in the spectral function,
\begin{equation}
A^{(1)}({\bf k},\omega) \sim (1-\tilde{\omega}^2)^{-1/2} 
\ln \left({(k_{x}-cos^{-1}
\tilde{\omega})^2+q_{D}^2\over{(k_{x}-cos^{-1}\tilde{\omega})^2}}\right).
\end{equation}
Here $q_{D}$ is the radius of the cylinder. For $\omega=\omega_{0}+ 2t cos
k_{x}$, the $x$-component of the phonon momentum is zero, and the
singular matrix element squared ($\sim 1/q^2$) integrated over $q_{y}$
and $q_{z}$ yields a singularity.

The 'long-range' dispersive  features appear  in the 2D single-phonon
spectral fuction as well. This function is reduced to
\begin{eqnarray}
A^{(1)}({\bf k},\omega) &\sim& \int _{a(\omega)}^{b(\omega)} dz
[1-(\tilde{\omega}-{1\over{4}}\cos z)^{2}]^{-1/2} \cr 
\times  [&\phi& (z,\omega,k_{x}, k_{y})+ \phi (z,\omega,k_{x},- k_{y}) \cr 
&+&\phi (z,\omega,-k_{x}, k_{y})+\phi (z,\omega,-k_{x}, -k_{y})],
\end{eqnarray}
with 
\begin{eqnarray}
\phi (z,\omega,k_{x},k_{y})=
[(x-k_{x})^2 +(z-k_{y})^2]^{-1/2} \cr \tan^{-1}\left({\pi\over{3}}[(x-k_{x})^2 +(z-k_{y})^2]^{-1/2}\right)
\end{eqnarray}
and $ x=\cos^{-1}(\tilde{\omega} -{1\over{4}}\cos z)$. The
integration limits are defined as $a(\omega)=
\cos^{-1}[4(\tilde{\omega}+1)]$ if $-5/4 \leq \tilde{\omega}\leq
-3/4$, $a(\omega)=0$ if $-3/4 \leq \tilde{\omega}\leq
5/4$, and $b(\omega)=
\cos^{-1}[4(\tilde{\omega}-1)]$ if $3/4 \leq \tilde{\omega}\leq
5/4$, $b(\omega)=\pi$ if $-5/4 \leq \tilde{\omega}\leq
3/4$. The numerical results are shown in Fig.2. 

We do not expect any {\it dispersive} peaks in the multiphonon ($l\geq 2$)
contributions to the spectral function because of  the additional
integrations compared with the single-phonon term. However, the 
nondispersive vHs remains for all $l\geq 1$. They can be washed out by the
phonon frequency dispersion in real crystals.

\vspace{0.5cm}

\centerline {\bf 4. Conclusions }

 We have calculated the Green's and spectral functions of a fermion on
a lattice coupled with  lattice vibrations via the long-range
Fr\"ohlich interaction. In a sharp contrast with the Holstein polaron 
the mass renormalisation exponent  of 
 Fr\"ohlich polaron ($Z'$) differs from the  renormalisation of the coherent
spectral weight ($Z$).  On the one hand this important result tells us that small
polarons
as well as   intersite bipolarons are perfectly mobile and can account for
the high-T$_{c}$ values in cuprates \cite{alecom}. On the other hand the
 coherent spectral weight remains strongly suppressed
in polaronic conductors, Eq.(24), because $Z$ might be less than $Z'$
 by one or even a few orders of magnitude. These unusual
 spectral features  provide an explanation for the apparent discrepancy
 between a very small Drude weight and a relatively moderate
  mass enhancement, $m^{*}\sim 3m_{e} - 10 m_{e}$  (depending on doping)
 of carriers in manganites \cite{cmr,des}
 and cuprates. They also explain why the  'extended' vHs observed in 
angle-resolved photoemission
 \cite{gof,mat} can be hardly seen in  angle- averaged
 photoemission. Indeed, the integrated spectral weight of the
 incoherent background is proportional to $(1-Z)$. It turns out to
be much
 larger than
 the coherent contribution, proportional to $Z<<1$. Finally, the ${\bf k}$
 dependent $incoherent$ background, Fig.1,2,  might obscure the experimental
 determination of a Fermi-level crossing. We believe that our
results can provide a quantitative approach to the experimental
tunneling and photoemission spectra. In particular the coherent part
of ARPES and tunneling  in
cuprates has been recently explained \cite{aleden}.

The authors greatly appreciate enlightening discussions with Chris Dent.
 C.S. has been supported in this
 work by a grant from the Royal Thai Government.

\centerline{{\bf Figure Captures}}

Fig.1. 1D single-phonon contribution to the polaron spectral function.

Fig.2. 2D single-phonon contribution to the polaron spectral function
along the $\Gamma-Y$ direction.


\begin{thebibliography}{99}
\bibitem{ref2}
Eq.(22) can be also derived from the polaronic Matsubara GF \cite{ale}.
\end{thebibliography}
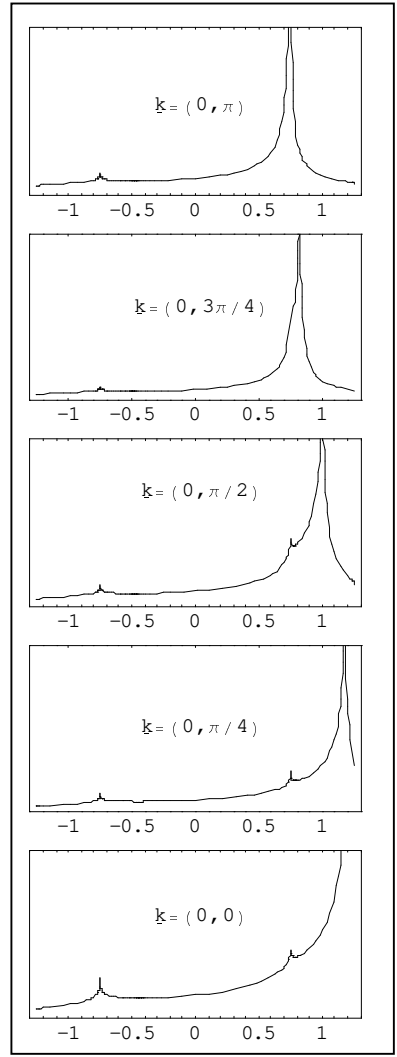
\end{document}